\documentclass[preprint,12pt]{elsarticle}

\def\rreq{{\it rreq}}
\def\rrep{{\it rrep}}

\def\unconfirmed{{\it unconfirmed}}
\def\valid{{\it valid}}

\newenvironment{example}{\paragraph{Example}}{\hfill$\square$}

\usepackage{amssymb}
\usepackage{graphicx}
\usepackage{caption}
\usepackage{subcaption}
\usepackage{url}
\usepackage[usenames,dvipsnames]{color}
\usepackage[usenamesdvipsnames]{xcolor}
\usepackage{mathtools}
\usepackage{perpage}
\usepackage{hyperref}
\usepackage{booktabs}
\usepackage{multicol}
\usepackage{float}
\usepackage{listings}
\usepackage{alltt}
\usepackage[figure,noline]{algorithm2e}
\usepackage{tikz}
\usepackage{bussproofs}

\lstdefinelanguage{rebeca}{
    morekeywords={reactiveclass, knownrebecs, statevars, main, msgsrv, constraints,con, main, define, LTL, CTL, boolean, int, shortint, byte, if, else, while, for, wait, msg, reset, set, self, false, true, now, after, delay, deadline, initial},
    otherkeywords={=>,<-,<\%,<:,>:,\#,@},
    sensitive=true,
    morecomment=[l]{//},
    morecomment=[n]{/*}{*/},
    morestring=[b]",
    morestring=[b]',
    morestring=[b]"""
}
\journal{Computer Networks}

\begin{document}

\begin{frontmatter}


\title{An Efficient Loop-free Version of AODVv2}

\author{Behnaz Yousefi}
\ead{b.yousefi@alumni.ut.ac.ir}
\author{Fatemeh Ghassemi\corref{cor1}}
\cortext[cor1]{Corresponding author}
\ead{fghassemi@ut.ac.ir}

\address{School of Electrical and Computer Engineering, College of Engineering,\\
University of Tehran, Tehran, Iran}

\begin{abstract}
Ad hoc On Demand distance Vector (AODV) routing protocol is one of the most prominent routing protocol used in Mobile Ad-hoc Networks (MANETs). Due to the mobility of nodes, there exists many revisions as scenarios leading to the loop formation were found.  We demonstrate the loop freedom property violation of AODVv2-11, AODVv2-13, and AODVv2-16 through counterexamples. 
We present our proposed version of AODVv2 precisely which not only ensures loop freedom but also improves the performance.
\end{abstract}

\begin{keyword}
Routing protocol \sep mobile Ad hoc network\sep
loop-freedom \sep AODV \sep modeling and verification.
\end{keyword}

\end{frontmatter}

\section{Introduction}\label{sec::Intro}
Mobile Ad-hoc Networks (MANETs) have different applications from military to disastrous situations where there is no network infrastructure and nodes can freely change their locations due to mobility of nodes. Mobility is the main feature of MANETs which makes them powerful and at the same time error prone in practice. The process of the protocol design are not straightforward and simulations are used to validate the protocol. However, all possible scenarios are not covered during simulations.

Since there is no base station or fixed network infrastructure, every node acts as a router and keeps the track of the previously seen packets to efficiently forward the received messages to desired destinations. In essence, MANETs need routing protocols in order to provide a way of communication between two indirectly-connected nodes. 

Ad hoc On Demand Distance Vector (AODV) routing protocol \cite{perkins1999ad} is one of the most popular routing protocol used in MANETs. It has two main versions, each one with several subversions. The AODV specification is given in plain English, no pseudo code or implementation is provided. It brings out lots of ambiguities which may lead to different implementations, or even worse, could cause the violation of important properties of AODV such as loop freedom. For example, while modeling the ADOV
v2 (version 11), we confronted some ambiguities that we worked them out through communication with the AODV authors. For instance, when there are more than one \emph{unconfirmed} routes, which one is going to be used? The answer was the best one as we speculated. Also what happens if it fails to receive an \emph{ack} from the best one? The answer was it is going to use the second best route and go on till it gets an \emph{ack}. It was also not clear when \textit{rerr} messages are sent while dealing with \emph{unconfirmed} routes, which turns out that they are never going to be sent if the route is \emph{unconfirmed}. 
Therefore, it is really necessary to have a precise specification while easy to read and understand. 
wRebeca modeling language was introduced in \cite{FOAC17} for the formal specification and verification of MANET protocols. It not only provides a means to specify a protocol precisely in a Java-like syntax, but also it is supported by a tool to verify given properties, e.g. loop freedom, on the protocol. Besides, adding new features or updating the existing network protocols invalidates all the verifications that have been done on the older versions. Therefore, as the process of developing the AODV protocol is an ongoing one, its verification should be too. In \cite{FOAC17}, the applicability of wRebeca is shown through the modeling and verification of the AODVv2 (version 11) protocol.

In this paper, we focus on several versions of AODVv2 and their shortcomings to assure loop freedom. First, we provide a short introduction of wRebeca and its important aspects in Section \ref{sec::wRebec}, and then we proceed by explaining AODVv2 concisely 
in Section \ref{sec::AODV}. We demonstrate the routing table maintenance procedure of its subversions and in their consequent, the scenarios leading to the violation of the loop freedom property of AODVv2-11, AODVv2-13, and AODVv2-16 through counterexamples in Section \ref{sec::loop}. Such scenarios, found automatically by our framework, have been communicated with the AODV group to be validated.
We explain the reason in the protocol design which leads to the loop freedom violation of AODVv2-16 and present two solutions to amend the protocol in Section \ref{sec::solutions}. Then, we discuss an excessive restriction, which the protocol applies to ensure its loop freedom, and its consequence on the performance of the protocol. 
Finally, we present our proposed version of AODVv2-16 which not only ensures loop freedom but also improves the performance. Inspecting all loop scenarios, makes it clear that loop formations are caused by updating the routing table not carefully enough. In fact, there are many factors that must be considered while updating the routing table, such as sequence numbers and route costs. Keeping the routing table loop-free even gets more sophisticated by maintaining more than one route per each destination. For example, in case only one route per each destination exists in the routing table, a new route with a greater sequence number than the existing ones simply replaces it, but now should it be added to the routing table or replace all other routes? As we see in Section \ref{sec::loop}, adding such a new route 
to the routing table may lead to the loop formation. As a matter of fact, the main cause of loop formation in AODVv2 (version 13) and (version 16) was mishandling the situation as a consequence of which a new route with the greater sequence number is added to the routing table. In the all versions of AODV, there is a function which enforces the loop freedom condition 
through verifying that each incoming route is not a sub-section of any existing routes. 
Nevertheless, when an incoming route has a greater destination sequence number, this function gets ignored to value its freshness. This kind of avoidance does not cause any problem when there is at most one route to each destination and the incoming route updates the existing one. However, when there are more than one route to each destination, the incoming route does not update all the existing ones which may lead to the loop formation.
    \section{Actor Model and the wRebeca Language}\label{sec::wRebec}
The computational model of actors \cite{agha1985actors,hewitt1977viewing} has been
introduced for the purpose of modeling concurrent and distributed
systems. Such modeling has become very popular in practice \cite{hewitt2008orgs,hewitt2009actorscript,karmani2009actor}\footnote{Scala programming language supports actor-models \url{http://www.scala-lang.org}}. Actors, the primitives of computation, are independent, well encapsulated, and of course run concurrently. Each actor has its own state indicated by its state variables and its encapsulation prohibits other actors to access its state variables directly. Each actor communicates with others only through message passing and owns a mailbox with a unique address to store the received messages. The behavior of an actor is defined in terms of a set of \emph{message server}s which specify how the actor reacts upon processing each received message. For example, if one actor wants to change the other actor state variable, it should do it through sending an update message to the other actor. The way this message is going to be processed is declared in the corresponding  message server of the other actor.  In this model,
message delivery is guaranteed but is not in-order. This policy
implicitly abstracts from the effects of network, i.e., delays over different routing paths, message conflicts, etc., and
consequently makes it a suitable modeling framework for concurrent
and distributed applications. 
The modeling language Rebeca \cite{SirjaniRebeca} provides
an operational interpretation of the actor model through a Java-like
syntax to fill the gap
between formal verification techniques and the real-world software
engineering of concurrent and distributed applications. It is empowered through various extensions introduced for different domains such as probabilistic systems \cite{varshosaz2012modeling}, real-time
systems \cite{TimedRebeca}, software product lines \cite{Sabouri}, and broadcasting environment \cite{bRebeca}. 

Mobility is the intrinsic characteristic of the MANETs which affects the correctness of MANET protocols. We extended Rebeca with the concepts of MANETs to model such networks in a more succinct way, so-called \textit{wRebeca} \cite{FOAC17}. It is supported by a toolset for efficient verification of wRebeca models 
regarding the mobility of nodes. 
wRebeca provides essential primitives for the modeling of MANET protocols, namely unicast, multicast and broadcast communications, abstracting the services of the data link layer. Furthermore, the concepts of connectivity and the underlying topology are considered for actors.  The message delivery is guaranteed for the receiving actors connected to a sender and also is in-order as communications are one-hop. 
Such an extension allows the modeler to setup the initial topology and specify the dynamic aspect of the networks, i.e., how the underlying topology changes through the novel concept of network constraints. A network constraint establishes a set of static (dis)connectivity relations among the nodes. Therefore, a wRebeca model is analyzed for all mobility scenarios respecting the constraints. The wRebeca is reasonably suitable for modeling MANET protocols. In this setting  each network node executing an instance of a MANET protocol can be represented through an actor with some state variables and message servers. There is a complete mapping between messages defined by the \emph{protocol specification}, e.g., an IETF draf, and message servers. The content of the message is passed through the message server arguments. The body of a message server encodes how a received message is going to be processed as defined by the specification. The information required to be maintained by each node is modeled by state variables. Hence, there is a good
traceability, from the model back to the protocol. If we find a problem in the model,
then we can trace it back into the protocol easier. The faithfulness of the framework to the MANET domain, make it usable for analysis and design of such protocols \cite{Marjan2018}.

As mentioned earlier, wRebeca is an extension of Rebeca with a Java-like syntax to easily read and apply. 
Every wRebeca model consists of two parts: the reactive class declaration part and the main part. Various components of the system are modeled through declaring different reactive classes.  Each reactive class has two major parts: one for maintaining its state, which is called \emph{statevars}, and the other for specifying its reaction upon receiving different messages, i.e., \emph{message server}s. 
The body of message servers may consist of conditional, assignment, and communication statements.  The syntax of communication statements that worth mentioning are broadcast, multicast, and unicast. Broadcasting a message is like calling a function, by indicating a message server name along with its parameters. Unicasting/Multicasting a message is slightly different since we need to mention the receivers. In addition, in case of unicast the  modeler can specify what is going to happen regarding to the success or failure of the communication. 

The second part of a wRebeca model is the \emph{main} part which declares the instances of defined reactive classes and their initialization. Furthermore, the modeler can define a set of constraints 
to restrict the topology changes in network. For instance, if it is known that two nodes would never get connected, i.e, they would never get into each other communication range, the topologies in which these node are connected together can be ruled out from possible topologies by expressing a constraint by which the link between these node is disconnected.

\begin{example}
\figurename{~\ref{code:aodv}} shows the wRebeca specification of the AODV protocol. َFor brevity some parts of the code have been abstracted away. Network nodes running an instance of AODV are modeled by a reactive class, lines (1-60). Each node has a routing table and an IP which is modeled by the state variables, lines (2-5). Every node can receive different routing messages, i.e. $\it rreq$, $\it rrep$, and $\it rerr$. The procedure of handling these messages is modeled through declaring different message servers while each one is responsible for handling a specific message. For example, the message server $\it rec\_rreq$  is responsible for handling received  $\it rreq$ messages, lines (21-41). In line 38, an $\it rreq$ message is broadcast while an $\it rrep$ is unicast in line 26.  Modeler has specified the behavior of the protocol based on the delivery status, lines (28-35). The second part of a Rebeca model is the \emph{main} part, lines (65-70), where rebecs get instantiated from declared reactive classes, for example $n1$ from $Node$. The first pair of parentheses specify the initial topology by indicating the name of other rebecs which are initially in the rebec neighborhood. For example, $n2$ is initially connected to $n1$ and $n4$. The second pair of parentheses, after colon, specify the parameters of the initial message which is going to be processed by the declared \emph{initial} message server. Each reactive class declaration at least has one message server namely \emph{initial} which acts like a constructor in object-oriented languages and used for initialization purposes, lines (6-9). For example, initializing state variables, routing table variables and starting a new route discovery by sending a new packet. As mentioned earlier, the main part in wRebeca has another part named network \emph{constraints}, lines (69-71). This part is used to reduce the domain of the possible topologies. For example, if it is  impossible for $n1$  to get out of the communication range of $n2$ and vice versa, modeler can express this situation by declaring a network constraint containing the relation ${\it con}(n1, n2)$. 
\end{example}
    \section{Overview of AODVv2 }\label{sec::AODV}
The AODV protocol is under continuous development and its working group publish a new version at most every $6$ months with the aim to improve the protocol and amend its shortcomings. However, all its (sub)versions almost follow the same design concept. More specifically, it uses some specific routing packets, e.g., $\it rreq$, $\it rrep$, and $\it rerr$, but the way these packets are sent and processed differs in every version. 
In this section, we briefly explain the common procedure of route discovery and maintenance among its variants. The wRebeca specification of its common code between versions 10, 11, 13, and 16 is given in \figurename{~\ref{code:aodv}}. Some parts of the code, abstracted in this specification, e.g., the one commented by ``processing code'', vary in different versions. 
\begin{figure}
	\begin{center}
		\begin{lstlisting}[language=rebeca,multicols=2]
reactiveclass Node(){
	statevars{
		int sn,ip;
		int[] dsn,rst,hops,nhop;
	}
	msgsrv initial(int i, 
		boolean starter){ 
		... /*Initilization code*/
	}			
	msgsrv rec_newpkt(int data,int dip_)
	{
		if(rst[dip_]==1) 
			{... /*forward packet*/}
		else {                       
			sn++;
			rec_rreq(0,dip_,
				dsn[dip_],self,sn,self,5);}
	}   		
	msgsrv rec_rreq (int hops_, int dip_ , int dsn_ , int oip_ , int osn_ , int sip_, int maxHop) 
	{    
		boolean gen_msg = false;
		... /*processing code*/
		if (gen_msg == true) {
			if (ip == dip_) {
				sn = sn+1;
				unicast(nhop[oip_],
					rec_rrep(0 , dip_ , sn , oip_ , self))
				succ:{
					rst[oip_] = 1;
					 }
				unsucc:{
					if(rst[oip_] == 1) 
						{... /*error*/}
					rst[oip_] = 2;}
			} else {
				hops_ = hops_ + 1;
				if(hops_<maxHop) {
					rec_rreq
						(hops_,dip_,dsn_,oip_,
						osn_,self,maxHop);} 
	}}}  		
	msgsrv rec_rrep(int hops_ ,int dip_ ,int dsn_ ,
		int oip_ ,int sip_){      
		boolean gen_msg = false;
		... /*processing code*/
		if(gen_msg == true){
			if(ip == oip_ ){
			... /*forward packet*/ }
			else {
				hops_= hops_+1;
				unicast(nhop[oip_],rec_rrep
					(hops_,dip_,dsn_,oip_,self))
				succ:{
					rst[oip_]=1;
					 }
				unsucc:{
					if(rst[oip_] == 1) 
						{...} /*error*/
					rst[oip_] = 2;}
	}}}
	msgsrv rec_rerr(int source_ ,
		int sip_, int[] rip_rsn) 
	{... /*error recovery code*/}
}
main{
	Node n1(n2,n4):(0,true);
	Node n2(n1,n4):(1,false);
	...
	constraints{
		and(con(n1,n2), con(n3,n4)) 
	}  
}
		\end{lstlisting}
	\end{center}
\vspace{-2mm} \caption{The AODV specification given in wRebeca \label{code:aodv}}
\end{figure} 

In this protocol, routes are built upon route discovery requests and maintained in nodes routing tables for further use. The routing table contains information about discovered routes and their status: The following information is maintained for each route:
\begin{itemize}
	\item{\it SeqNum}: destination sequence number
	\item{\it route\_state}: the state of the route to the destination 
	\item {\it Metric}: indicates the cost or quality of the route, e.g., hop count, the length of the path from the node to the destination via the respective next hop
	\item {\it NextHop}: IP address of the next hop to the destination
\end{itemize}
The routing table of each node is modeled by a set of variables of array type, namely ${\it dsn}$, ${\it rst}$, ${\it hops}$, and ${\it nhop}$ to denote {\it SeqNum}, {\it route\_state}, {\it Metric}, and {\it NextHop}, respectively. In addition, ${\it dip\_}$ and  ${\it oip\_}$ denote destination and originator IPs which are used as indexes to retrieve the information of a route to destination/originator in such arrays. For instance, ${\it rst}[oip\_]$ denotes the route state to the originator.

Whenever a node intends to send a data packet to another, i.e., when it receives a $\it newpkt$ message, it looks up its routing table to see if it has a valid route to the intended destination, line 12 of $\it rec\_newpkt$. In case it finds a route, it sends the data packet through the next hop specified in that route, otherwise it starts a route discovery by broadcasting a route request, i.e. $\it rreq$ after increasing its sequence number, lines (14-17) of $\it rec\_newpkt$. Whenever a node receives a new routing packet, $\it rreq$, it updates its routing table with new information to keep it up-to-date, abstracted code at line 22 of $\it rec\_rreq$. $\it rreq$ messages contain route towards a \textit{source} while $\it rrep$ messages carry route information towards a \textit{destination}. Therefore, as an $\it rreq$ packet proceeds towards the destination, in each node, a \emph{backward path}, a path to the \textit{source} from the node, gets constructed. Similarly, a \emph{forward path}, a path to the \textit{destination} from the node, is built while $\it rrep$ packets traverse the constructed \emph{backward path} from the \textit{destination} towards the \textit{source}.  
Each node upon receiving an $\it rreq$ message looks up its routing table and if it has a route to the requested destination it would reply through sending an $\it rrep$, lines (25-35) of $\it rec\_rreq$ otherwise, it resends the $\it rreq$ message after increasing the hop count if the maximum number of hop count limit is not reached, lines (36-40) of $\it rec\_rreq$. Whenever a node receives an $\it rrep$ message, it updates its routing table accordingly to construct a \emph{forward path}, the abstracted code at line 45 of $\it rec\_rrep$. When the $\it rrep$ reaches the source, abstracted code at line 48 of $\it rec\_rrep$, a bidirectional route has been formed and the data packet can be sent through next-hops on nodes routing tables towards the destination. When a node which is not the source receives a $\it rec\_rrep$ message, it unicasts the $\it rec\_rrep$ toward the source after increasing the hop count, lines (49-58) of $\it rec\_rrep$.  In addition to the $\it rreq$ and $\it rrep$ packets, there is an $\it rerr$ packet which is sent whenever a node fails to send a packet through a \emph{valid} route, line 33 of $\it rec\_rreq$ and line 58 of $\it rec\_rrep$, in order to informs other interested nodes in the broken route about the failure.

From version 10, a new ability has been added to the protocol to maintain more than one route to a destination. For each destination, multiple routes may exist with different next-hops, i.e., \emph{unconfirmed} next-hop, a next-hop which its bidirectionality has not been confirmed yet. Whenever an $\it rrep$ is going to send a package to an \emph{unconfirmed} next-hop, it must request an $\it ack$ from the receiver to become sure about its bidirectionality. This new feature improves the performance since for sending a packet there is no need to wait for a next-hop to get \emph{confirmed}, and consequently its route to become \emph{valid}. Although having multiple routes to one destination has its benefits, it can lead to a loop formation when it is used with not required precautions as we are going to explain in the following section.

\section{Loop formation Scenarios}\label{sec::loop}
We explain how different versions of AODVv2 protocol try to prevent loop formation and how they fail to do so through counterexamples which are obtained by our tool. The AODV protocol manuscript has different sections, e.g. initialization, adjacency monitoring, route maintenance and processing received route information \cite{perkins1999ad}. For the purpose of loop formation avoidance, we will focus on the \emph{processing received route information} part of the specification since a loop is formed if and only if preventative measures have not been taken to account while updating the routing tables. Therefore, in the section for the sake of simplicity, we only focus on evaluating received route information and consequently updating the routing tables, abstracted in the specification of \figurename{~\ref{code:aodv}} and commented by ``processing code''. For a comprehensive specification, we refer the interested reader to their corresponding IETF drafts.

\subsection{AODVv2-11}
This version maintains more than one next hop per each destination which increases the probability of packet delivery since if one route gets broken, there may be other routes that can be used as an alternative.\footnote{\url{https://tools.ietf.org/html/draft-ietf-manet-aodvv2-11}} When there are more than one route, the best one would be used. The best route is chosen based on the concept of \emph{route state} and \emph{cost}. Route states are determined by the concept of \emph{neighbor state}s of next hops which determines the adjacency states of the node's neighbors, and can have one of the following values: 
\begin{itemize}
\item \it{Confirmed}: indicates that a bidirectional link to that neighbor exists. This state is achieved either through receiving an \it{rrep} message in response to a previously sent \it{rreq} message, or an \it{rrep\_ack}  message as a response to a previously sent \it{rrep} message (requested an \it{rrep\_ack}) to that neighbor.
\item \it{Unknown}: indicates that the link to that neighbor is currently unknown. Initially, the states of the links to the neighbors are unknown.
\item \it{Blacklisted}: indicates that the link to that neighbor is unidirectional. When a node has failed to receive the \it{rrep\_ack} message in response to its \it{rreq} message to that neighbour, the  neighbor state is changed to blacklisted. Hence, it stops forwarding any message to it for an amount of time, \it{ResetTime}. After reaching the \it{ResetTime}, the neighbor's state will be set to \it{unknown}.
\end{itemize}
Such information are  kept in the neighbor table of each node. Route states, the states of the routes to each destination, are kept in the routing table and can have one of the following values:
\begin{itemize}
\item \emph{unconfirmed}: when the neighbor state of the next hop is unknown;
\item \emph{active}: when the link to the next hop has been confirmed, and the route is currently used;
 \item \emph{idle}: when the link to the next hop has been confirmed, but it has not been used in the last \it{active\_interval};
\item \emph{invalid}: when the link to the next hop is broken, i.e., the neighbor state of the next hop is blacklisted.
\end{itemize}
A route is called \emph{valid} if it is either active or idle. Although there can exist more than one \emph{unconfirmed} route to each destination, there can be only one \emph{valid} route to each destination. When a route state to a destination gets changed to \emph{valid}, all the routes to the same destination are removed from routing table.

\subsubsection{Updating the Routing Table}\label{subsec::update}
Every received route message contains a route and consequently is evaluated to check for any improvement. Note that an $\it rreq$ message contains a route to its source while an $\it rrep$ message contains a route to its destination. Therefore, as the routes are identified by their destinations
, in the former case, the destination of the route is the originator of the message 
and in the latter, it is the destination of the message. 
Note that we say a router is \emph{better} then others if it has either a greater sequence number than others or an equal sequence number while its cost, e.g., hop count, is less than others. The routing table must be updated if one of the following conditions is realized:
\begin{itemize}
	\item no route to the destination exists in the routing table:  the route is added to the routing table.
	\item all the existing routes to the destination are \emph{unconfirmed}, i.e., their next hops are \emph{unconfirmed}: the route is added to the routing table.\label{case::unconfirmed}
	\item the incoming route is a better route than the  existing valid one: 
	if the next hop of the incoming route is \emph{confirmed}, it updates the existing valid route with the received route, otherwise it adds the received route to the routing table since it may be confirmed in the future and consequently, replaces the existing route.
	\item the incoming route is a better route than the  existing invalid one:
	it updates the existing invalid route with the incoming route.
\end{itemize}

\subsubsection{Loop Formation Scenario}\label{subsec::loop11}
In this version no constrain has been applied to the \emph{unconfirmed} next-hop of an incoming route prior its addition to the routing table when the route status of the existing routes are \emph{unconfirmed} (the second case in Section \ref{subsec::update}). This lack of restriction easily leads to a looping scenario which is described in the following. Assume that each route entry of the routing table has the following format: $({\it dest}, {\it next\_hop},{\it hop\_count}, {\it seq\_num}, {\it route\_state})$, where the first element indicates IP of the destination, the second, IP of the next hop, the third, the length of the route to the destination, the forth, the sequence number of the destination, and the last, the route state, respectively.  
Consider a network of four nodes as shown in \figurename{~\ref{Fig::Net}.

\begin{figure}
	\centering
\begin{minipage}{.45\linewidth}
	\centering
	\begin{tikzpicture}[scale=.7, transform shape]
	\node[style=circle,draw] (n3) at (1,1) {$n_3$};
	\node[style=circle,draw] (n1) at (1,3) {$n_1$};
	\node[style=circle,draw] (n4) at (0,2) {$n_4$};
	\node[style=circle,draw] (n2) at (2,2) {$n_2$}; 
	\draw (n3)
	edge (n4); \draw (n3) edge (n2); \draw (n1) edge (n4); \draw (n2)
	edge (n1); \draw (n4) edge (n2);
	\end{tikzpicture}    
	\caption{The network topology}
	\label{Fig::Net}
\end{minipage}%
\hfill
\begin{minipage}{.55\linewidth}%
	\centering
	\begin{subfigure}[b]{0.34\textwidth}
		\begin{tikzpicture}[scale=.7, transform shape]
		\node[style=circle,draw] (n4) at (1,1) {$n_4$};
		\node[style=circle,draw] (n1) at (1,3) {$n_1$};
		\node[style=circle,draw] (n2) at (0,2) {$n_2$};
		\node[style=circle,draw] (n3) at (2,2) {$n_3$};
		\draw (n3)
		edge (n2); \draw (n3) edge (n4); \draw (n3) edge (n1); \draw (n2)
		edge (n4); 
		\end{tikzpicture}  
		\caption{}
		\label{Fig::Netv13-1}
	\end{subfigure}
	\begin{subfigure}[b]{0.3\textwidth}
		\begin{tikzpicture}[scale=.7, transform shape]
		\node[style=circle,draw] (n4) at (1,1) {$n_4$};
		\node[style=circle,draw] (n1) at (1,3) {$n_1$};
		\node[style=circle,draw] (n2) at (0,2) {$n_2$};
		\node[style=circle,draw] (n3) at (2,2) {$n_3$};
		\draw (n3)
		edge (n2); \draw (n3) edge (n4); \draw (n2) edge (n1); \draw (n2)
		edge (n4); 
		\end{tikzpicture}    \caption{}
		\label{Fig::Netv13-2}
	\end{subfigure}			
	\caption{Two possible network topologies for a network of four nodes}
	\label{Fig::Netv13}

\end{minipage}%
\end{figure}
\begin{enumerate}
	\item
	$n_1$ initiates a route discovery procedure for destination $n_3$ by broadcasting an $\rreq$ message with the sequence number $2$.
	\item $n_2$ receives $\rreq$ message as it is a neighbor of $n_1$. Since it is the first time that $n_2$ has received an $\rreq$ message from $n_1$, the neighbor state of $n_1$ is set to \emph{unconfirmed}. Therefore, the route state of the received route is \emph{unconfirmed}, and $n_1$ adds  the incoming route $(n_1, n_1, 1, 2, \unconfirmed)$ to its routing table. Since $n_2$ is not the intended destination of the route request, it rebroadcasts an $\rreq$ message.
	\item $n_4$ also receives the $\rreq$ message sent by $n_1$ (simultaneous with $n_2$) and inserts the incoming route $(n_1, n_1, 1, 2, \unconfirmed)$ to its routing table towards $n_1$ similar to $n_2$. Then, it rebroadcasts the $\rreq$ message.
	\item $n_2$ after receiving the $\rreq$ message sent by $n_4$, adds the route $(n_1, n_4, 2, 2,\\ \unconfirmed)$ to its routing table since the existing route to $n_1$, i.e., $(n_1, n_1, 1, 2, \unconfirmed)$, is \emph{unconfirmed}.
	\item $n_4$ also adds $(n_1, n_2, 2, 2, \unconfirmed)$ to its routing table after processing the message $\rreq$ sent by $n_2$. At this point a loop is formed between $n_2$ and $n_4$. \label{case::loopunconfiremd}
	\item $n_3$ receives  the $\rreq$ message sent by $n_1$ via $n_2$, and since it is the destination, it sends an $\rrep$ message towards $n_2$.\label{case::loop1con}
	\item Assume that $n_1$ moves out of the communication ranges of $n_2$ and $n_4$.
	\item
	$n_2$ receives the message $\rrep$ sent by $n_3$ and as the
	route state of the routes towards $n_1$ is \emph{unconfirmed}, it unicasts
	an $\rrep$ message one by one to the existing next hops, i.e., $n_1$ and $n_4$, till it gets an ack. Due to the movement of $n_1$, it receives no ack from $n_1$ and the route with the next hop $n_1$ is removed from the routing table. However, it receives an ack from $n_4$. Therefore, the neighbor
	state of $n_4$ is set to \emph{confirmed} and subsequently the respective route
	state towards $n_1$ to \emph{valid}. 
	\item
	$n_4$ by receiving the message $\rrep$ from $n_2$ unicasts it
	to its next hops, i.e., $n_1$ and $n_2$,  similar to
	$n_2$. Since it only receives an ack from $n_2$, it updates its routing table by validating $n_2$ as its
	next hop to $n_1$, and hence a loop is formed between $n_2$ and $n_4$ over valid routes. \label{case::loop11end}
\end{enumerate}

\subsection{AODVv2-13}
After communicating our result on AODVv2-11 to the AODV group, they revised the protocol to restrict the addition of \emph{unconfirmed} routes when all the existing routes to a destination are \emph{unconfirmed}. Hence, only the second step of the procedure of Section \ref{subsec::update} is revised: an incoming route is added to the routing table if all the existing routes to its destination are \emph{unconfirmed} while the incoming is better the existing ones.\footnote{\url{https://tools.ietf.org/html/draft-ietf-manet-aodvv2-13}}

\subsubsection{Loop Formation Scenario}
Although the scenario of Section \ref{subsec::loop11} is prohibited, a loop scenario occurs due to resending the $\rreq$ messages in a network of four nodes with the topologies shown in \figurename{~\ref{Fig::Netv13}}. At first nodes are connected to each other as shown in \figurename{~\ref{Fig::Netv13-1}}.

 \begin{enumerate}
\item $n_1$ initiates a route discovery procedure for destination $n_4$ by broadcasting an $\rreq$ message to $n_3$ with the sequence number $2$.
\item $n_3$ inserts the incoming route $(n_1, n_1, 1, 2, \unconfirmed)$ to its routing table and broadcasts an $\rreq$ message to its neighbors, $n_2$ and $n_4$.
\item $n_2$ upon receiving the message $\rreq$  sent by $n_3$ updates its routing table and adds the incoming route $(n_1, n_3, 2, 2, \unconfirmed)$ to its routing table.
\item topology changes at this point and $n_2$ moves into the communication range of $n_1$, gets connected to $n_1$, while $n_3$ leaves the communication range of $n_1$, gets disconnected from $n_1$, which leads to the network topology shown in \figurename{~\ref{Fig::Netv13-2}}.
\item $n_1$, which has not received an $\rrep$ message yet, resends the message $\rreq$ after increasing its sequence number to $3$ (due to the timeout to receive such a reply).
\item $n_2$ receives the incoming route $(n_1, n_1, 1, 3, \unconfirmed)$, since it is a better route it would be added to the routing table. Then, $n_2$ broadcasts an $\rreq$ message to its neighbors, i.e., $n_3$ and $n_4$.
\item $n_3$ evaluates the received message sent by $n_2$ and adds the incoming route $(n_1, n_2, 2, 3, \unconfirmed)$ to its routing table since the sequence number of the received message is greater than the stored one, i.e, $(n_1, n_1, 1, 2, \unconfirmed)$. At this point a loop has been formed between nodes $n_2$ and $n_3$, similar to the step \ref{case::loopunconfiremd} of the scenario explained in Section \ref{subsec::loop11} for version 11. Therefore, continuing with a scenario similar to the steps \ref{case::loop1con}-\ref{case::loop11end} of the scenario for version 11, a loop is formed between $n_2$ and $n_3$ over valid routes. 
  \end{enumerate}
  
This loop scenario occurs because the existing unconfirmed route $(n_1, n_1, 1,\\ 2, \unconfirmed)$ has not been replaced by the received better route $(n_1, n_2, 2, 3,\\ \unconfirmed)$. Instead, the received new route is only added to the table. We remark that a new route replaces an existing one only when the route state of the existing route is  \emph{invalid} or the route state of the new route is \emph{confirmed}. 
 

\subsection{AODVv2-16} 
It is the last AODVv2 protocol which applies even more restrictions for updating the routing table to ensure loop freedom.\footnote{\url{https://tools.ietf.org/html/draft-ietf-manet-aodvv2-16}} It maintains at most two routes for each destination while one is \emph{(in)valid} and the other is \emph{unconfirmed}. To prevent loops in this version, an incoming route updates the existing route with the same status. In case no route exists with the same status, it will be added to the table. Therefore, the routing table always keeps  better routes for each status.

\subsubsection{Updating the Routing Table}
The updating procedure has been revised accordingly: 
 \begin{itemize}
 	\item no route exists to the destination: the route is added to the routing table.
 	\item the incoming route is better than the existing one. Two cases can be distinguished: 
 	\begin{enumerate}
 		\item there is only one matching route with the same destination: 
 		\begin{itemize}
 			\item the route state of the existing route is invalid: the incoming route must replace the existing one;
 			\item the route state of the incoming route and the existing one are the same: the incoming route should replace the existing one. 
 			\item the route state of the incoming route is \emph{unconfirmed} and it offers improvement to the existing \emph{valid} route: the incoming route should be added to the routing table.
 		\end{itemize}
 		\item there are two matching routes with the same destination where one is \emph{valid/invalid} and the other is \emph{unconfirmed}:
		\begin{itemize}
			\item if the incoming route offers improvement to the existing  route with the same status, then it should replace it.
			\item if the existing route is \emph{invalid} and the incoming route is \emph{valid}: the existing route is replaced by the incoming route even if the incoming route does not offer improvement.
		\end{itemize}
 	\end{enumerate}
 \end{itemize}
  
\subsubsection{Loop Formation Scenario}
The loop scenario is given for a network of four nodes with the network topologies shown in \figurename{~\ref{Fig::Netv16}} with the initial topology illustrated in \figurename{~\ref{Fig::Netv16-1}}: 
\begin{enumerate}
 	\item $n_1$ initiates a route discovery procedure for destination $n_4$ by broadcasting an $\rreq$ message to $n_3$ with the sequence number $2$.
 	\item $n_3$ inserts the incoming route $(n_1, n_1, 1, 2, \unconfirmed)$ in its routing table and broadcasts an $\rreq$ message to $n_2$ and $n_4$.
 	\item $n_2$ receives the message $\rreq$  sent by $n_3$ and updates its routing table by inserting the route $(n_1, n_3, 2, 2, \unconfirmed)$ into its routing table.
 	\item $n_2$ becomes aware that its connectivity to $n_3$ is bidirectional, for example through receiving an $\rrep\_{\it ack}$ from $n_3$ in response of a sent $\rrep$ message for another route, therefore the neighbor state of $n_3$ is updated to \emph{confirmed} and route states of all those routes which use $n_3$ as their next hops must be updated to \emph{valid}. As a result, the route entry $(n_1, n_3, 2, 2, \unconfirmed)$ of $n_2$'s routing table gets updated  to $(n_1, n_3, 2, 2, {\it valid})$.	
 	\item the topology changes at this point and $n_3$ moves out of the communication range of $n_1$ while $n_2$ enters the  communication range of $n_1$, which lead to the network topology shown in \figurename{~\ref{Fig::Netv16-2}}.
  	\item $n_1$ resends another $\rreq$ message with the increased sequence number of $3$ to $n_2$ (due to the timeout for receiving a reply).	
 	\item $n_2$ processes the received $\rreq$ message from $n_1$, since it has the greater sequence number than the stored one, it is used to update the routing table. As the stored route with next hop $n_3$ is \emph{valid}, the incoming route $(n_1, n_1, 1, 3, \unconfirmed)$ is added to the routing table as a new route. Then, $n_2$ broadcasts the received $\rreq$ to its neighbors. 
 	 \item the topology changes at this point and $n_3$ moves into the communication range of $n_1$, gets connected to $n_1$ which leads to network topology shown in \figurename{~\ref{Fig::Netv16-3}}. 
  	\item assume that the connectivity status of $n_3$ to $n_1$ becomes bidirectional, therefore the route entry $(n_1, n_1, 1, 2, \unconfirmed)$ of $n_3$'s routing table gets updated  to $(n_1, n_1, 1, 2, {\it valid})$.	 
 	\item $n_3$ receives the  incoming route $(n_1, n_2, 2, 3, \unconfirmed)$ via the  $\rreq$ message sent by $n_2$. Since the incoming route has a greater sequence number than the stored one and the stored one is \emph{valid}, it will be added to the routing table. At this point a loop between $n_2$ and $n_3$ is formed. Again by continuing with a scenario similar to the steps \ref{case::loop1con}-\ref{case::loop11end} of the scenario for version 11, a loop is formed between $n_2$ and $n_3$ over valid routes. 
 \end{enumerate}
 
 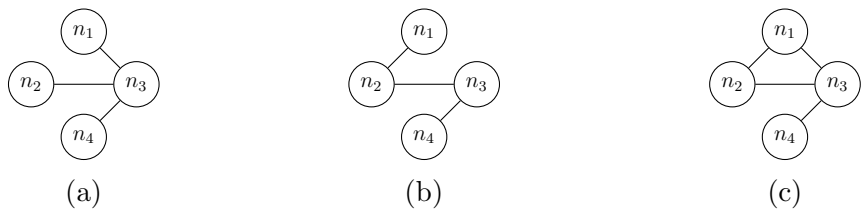
\begin{figure}
 	\centering
 	\begin{subfigure}[b]{0.3\textwidth}
 		\centering
 		\begin{tikzpicture}[scale=.7, transform shape]
 		\node[style=circle,draw] (n4) at (1,1) {$n_4$};
 		\node[style=circle,draw] (n1) at (1,3) {$n_1$};
 		\node[style=circle,draw] (n2) at (0,2) {$n_2$};
 		\node[style=circle,draw] (n3) at (2,2) {$n_3$};
 		\draw (n3)
 		edge (n1); \draw (n3) edge (n4); \draw (n3) edge (n2);
 		\end{tikzpicture}    \caption{}
 		\label{Fig::Netv16-1}
 	\end{subfigure}
 	\begin{subfigure}[b]{0.34\textwidth}
 	\centering
 		\begin{tikzpicture}[scale=.7, transform shape]
 		\node[style=circle,draw] (n4) at (1,1) {$n_4$};
 		\node[style=circle,draw] (n1) at (1,3) {$n_1$};
 		\node[style=circle,draw] (n2) at (0,2) {$n_2$};
 		\node[style=circle,draw] (n3) at (2,2) {$n_3$};
 		\draw (n2)
 		edge (n1); \draw (n3) edge (n4); \draw (n3) edge (n2);
 		\end{tikzpicture}  
 		\caption{}
 		\label{Fig::Netv16-2}
 	\end{subfigure}
 		\begin{subfigure}[b]{0.34\textwidth}
 		\centering
 			\begin{tikzpicture}[scale=.7, transform shape]
 			\node[style=circle,draw] (n4) at (1,1) {$n_4$};
 			\node[style=circle,draw] (n1) at (1,3) {$n_1$};
 			\node[style=circle,draw] (n2) at (0,2) {$n_2$};
 			\node[style=circle,draw] (n3) at (2,2) {$n_3$};
 			\draw (n2)
 			edge (n1); \draw (n3) edge (n4); \draw (n3) edge (n2);\draw (n3) edge (n1);
 			\end{tikzpicture}  
 			\caption{}
 			\label{Fig::Netv16-3}
 		\end{subfigure}
 	\caption{Three possible network topologies for a network of four nodes}
 	\label{Fig::Netv16}
 \end{figure}
 
By examining the counter example, we realize that a loop is formed as the loop freedom condition is not always considered and consequently, the new route will be added to the routing table when the sequence number is greater than the existing valid one. 
To amend this situation, we propose two options:
\begin{enumerate}
\item The loop freedom condition should be always considered. 
Therefore, if the new route does not satisfy the loop freedom condition, it must not be used to update the routing table even if it has a greater sequence number.
\item The new route with a greater sequence number will be added to the routing table while all the existing routes are removed from the routing table. 
\end{enumerate}  
These two approaches differs regarding how they prioritize a new route with a greater sequence number and an existing route. The first solution prefers to keep the valid one by ignoring the new route with a greater sequence number while the second one favors the new route with a greater sequence number over existing routes even the valid ones.
Nevertheless, we believe that there is a better approach which not only ensures loop freedom but also boosts the performance by maintaining more eligible routes for forwarding a packet to a destination. Irrespective of which solution is being adopted, we demonstrate through an example how the protocol fails to forward a packet while there could have existed a route if the routing table had been updated better. The example is given for a network which consists of seven nodes with the topologies shown in  the \figurename{~\ref{Fig::Netv16-nonefficiency}}.
\begin{figure}
	\centering
	\begin{subfigure}[b]{0.45\textwidth}
		\centering
		\begin{tikzpicture}[scale=.7, transform shape]
		\node[style=circle,draw] (n4) at (2,3) {$n_4$};
		\node[style=circle,draw] (n1) at (1,5) {$n_1$};
		\node[style=circle,draw] (n2) at (0,4) {$n_2$};
		\node[style=circle,draw] (n3) at (1,3) {$n_3$};
		\node[style=circle,draw] (n5) at (0,3) {$n_5$};
		\node[style=circle,draw] (n6) at (1,2) {$n_6$};
		\node[style=circle,draw] (n7) at (1,1) {$n_7$};
		\draw (n2) edge (n1); \draw (n1) edge (n3);\draw (n1) edge (n4);
		\draw (n2) edge (n5);
		\draw (n5) edge (n6); \draw (n3) edge (n6);\draw (n4) edge (n6);
		 \draw (n6) edge (n7);
		\end{tikzpicture}  
		\caption{}
		\label{Fig::Netv16-nonefficiency1}
	\end{subfigure}
	\begin{subfigure}[b]{0.45\textwidth}
		\centering
		\begin{tikzpicture}[scale=.7, transform shape]
		\node[style=circle,draw] (n4) at (2,3) {$n_4$};
		\node[style=circle,draw] (n1) at (1,5) {$n_1$};
		\node[style=circle,draw] (n2) at (0,4) {$n_2$};
		\node[style=circle,draw] (n3) at (1,3) {$n_3$};
		\node[style=circle,draw] (n5) at (0,3) {$n_5$};
		\node[style=circle,draw] (n6) at (1,2) {$n_6$};
		\node[style=circle,draw] (n7) at (1,1) {$n_7$};
		\draw (n2) edge (n1); \draw (n1) edge (n3);\draw (n1) edge (n4);
		\draw (n2) edge (n5);
		\draw (n4) edge (n6);
		\draw (n6) edge (n7);
		\end{tikzpicture}  
		\caption{}
		\label{Fig::Netv16-nonefficiency2}
	\end{subfigure}
	\caption{Two possible network topologies for a network of seven nodes}
	\label{Fig::Netv16-nonefficiency}
\end{figure}
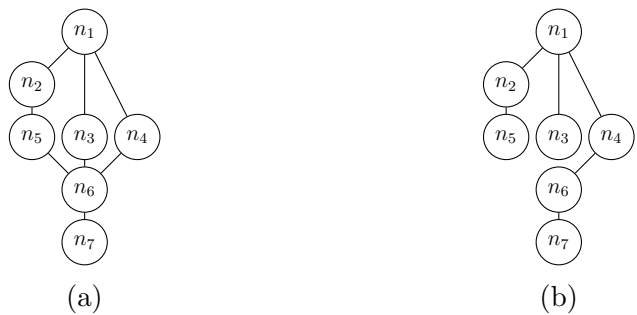
\begin{enumerate}
 	\item $n_1$ initiates a route discovery procedure for the destination $n_7$ by broadcasting an $\rreq$ message with the sequence number $2$.
 	\item $n_2$, $n_3$, and $n_4$  update their routing tables upon receiving the $\rreq$ message sent by $n_1$, and broadcast the $\rreq$ message to their neighbors.
 	\item $n_5$ receives the $\rreq$ message sent by $n_2$ and after updating its routing table broadcasts it. 
 	\item $n_6$ receives the $\rreq$ message sent by $n_5$ and adds the route $(n_1, n_5, 3, 2, \unconfirmed)$ to its routing table. Then, it broadcasts the $\rreq$ message to $n_7$.
 	\item assume that the connectivity status of $n_5$ to $n_6$ becomes bidirectional, therefore, the route $(n_1, n_5, 3, 2, \unconfirmed)$ gets updated to $(n_1, n_5, 3, 2, \valid)$.	
 	\item $n_6$ receives the $\rreq$ message sent by $n_3$ and  since it is a better route and the stored one is a \emph{valid} one, the incoming route $(n_1, n_3, 2, 2, \unconfirmed)$ is added to the routing table.	
 	\item $n_6$ receives the $\rreq$ message sent by $n_4$ and  since it doe not improve the existing \emph{unconfirmed} route, it gets discarded. 
 	\item the topology changes at this point as $n_5$ and $n_3$ move out of the communication range of $n_6$ which leads to the network topology shown in \figurename{~\ref{Fig::Netv16-nonefficiency2}}. 	
 	\item $n_7$ receives the $\rreq$ message sent by $n_6$ and since it is the destination, it replies through sending an $\rrep$ message to $n_6$.	 
 	\item $n_6$ receives the $\rrep$ message sent by $n_7$. To forward its $\rrep$ message to the originator, i.e. $n_1$, it has two next hops in its routing table to the destination $n_1$: $n_3$ and $n_5$. Both next hops are going to fail to deliver the message since they have got disconnected due to the topology change. Although the route through $n_4$ does exist, it had been ignored. 
 \end{enumerate}
As the number of nodes increases, the chance of having more than one \emph{unconfirmed} route and consequently, the effect of ignoring them on the performance raises. In the following section, we present a solution which not only satisfies the loop freedom invariant but also improves the performance by preserving multiple routes for each destination.  

\section{Proposed Procedure for Updating the Routing Table}\label{sec::solutions}
According to the scenarios mentioned for the different versions of AODV, the main reason leading to loop formation is ignoring the
loop freedom condition. In the previous subsection we presented two solutions. While these solutions are loop-free, they impose some restrictions which can degrade the performance. Hence, we present the modified version of these approaches while lifting the unnecessary restriction so that it is possible to have multiple routes to the same destination. Although it is possible to have an infinite number of routes, it is more realistic to bound it since there is a trade off between the storage cost and the performance.   

\subsection{Solution 1: Preferring Hop count to  Freshness} \label{subsec::Solution1}
In this approach we treat an incoming route with a greater sequence number in a same way we handle an incoming route with an equal sequence number compared to the existing routes. It means that the loop freedom condition is always checked. The procedure of evaluating the incoming route and updating the routing table is modified accordingly:

\begin{figure}
	\begin{center}
		\begin{lstlisting}[language=rebeca, multicols=2]
if(dsn[oip_][0]==-1)
{
	dsn[oip_][0]=osn_;
	if(neigh_state[sip_]==true){
		rst[oip_][0]=1;
	} else 	{
		rst[oip_][0]=0; }
	hops[oip_][0]=hops_;
	nhop[oip_][0]=sip_;
	dsn[oip_][0]=osn_;
	
	process_msg = true;
}else{
	boolean loopFree=true;
	for(int i=0;i<N;i++){
		if(dsn[oip_][i] == -1) {
			continue;
		}
		if(dsn[oip_][i] > osn_ || hops[oip_][i] < hops_){
			loopFree=false;
			break;
		}
	}
	if(loopFree){
		for(int i=0;i<N;i++){
			if(nhop[oip_][i]==-1 || nhop[oip_][i]==sip_)
			{
				route_num = i;
				break;
			}
		}
		hops[oip_][route_num]=hops_;
		nhop[oip_][route_num]=sip_;
		dsn[oip_][route_num]=dsn_;
		if(neigh_state[sip_]==true)
		{
			rst[oip_][route_num]=1;
			for(int i=0;i<N;i++)
			{
				dsn[oip_][i]=-1;
				hops[oip_][i]=-1;
				nhop[oip_][i]=-1;    
			} 
		} else {
			rst[oip_][route_num]=0;
		}
		process_msg = true;
	}			
}		
		\end{lstlisting}
	\end{center}
	\vspace{-2mm}
	\caption{The first solution for updating the routing table
		\label{code:UpdatingRoutingTable-solution1}}
\end{figure}
\begin{itemize}
	\item no route exists to the destination: the route is added to the routing table.
	\item the sequence number of the incoming route is equal or greater than all the existing routes to the same destination while its cost, e.g., hop count, is equal or less than all the existing routes: the incoming route is added if the bound has not been already reached.

\item otherwise, no change is applied to the table. 
\end{itemize}

The precise specification of this procedure is depicted in the  \figurename{~\ref{code:UpdatingRoutingTable-solution1}}. This code replaces the abstracted code at line 22 in the specification of \figurename{~\ref{code:aodv}} in the body of the message server handling $\rreq$. We remark that $\rreq$ and $\rrep$ messages have parametrizes such as the hop count that the message has been relayed from the originator, destination IP, destination sequence number, originator IP (the origin of the message), and sender IP, specified by $\it hops\_$, $\it dip\_$, $\it dsn\_$, $\it oip\_$, and $\it sip\_$, respectively. As the destination of the route in an $\rreq$ message is its originator, this code uses ${\it oip\_}$ in its evaluations. However, the destination of the route in an $\rrep$ message is identified by ${\it dip\_}$. Therefore, the code replacing the abstracted code at line 45 in the body of the message server handling $\rrep$ will be the same while ${\it dip\_}$ is used in the evaluations. 
As more than one route is maintained for each destination, the variables $\it dsn$, $\it rst$, $\it hops$, $\it nhop$ of the specification of \figurename{~\ref{code:aodv}} become two dimensional of size $n\times n$, where $n$ is the number of nodes. The second dimension keeps information of the alternative routes via different next hops for each destination. Thus, ${\it hops}[i][j]$ indicates the hop count of the $j$-th route to the destination $i$. Similarly,  ${\it rst}[i][j]$ refers to the state of the $j$-th route to the destination with IP $i$ which can have the values $0$, $1$, or $2$ to indicate that the route is \emph{unconfirmed}, \emph{valid}, or \emph{invalid}, respectively. Furthermore, a variable ${\it neigh\_state}$ is added to the specification to keep the adjacency state of the neighbors, where ${\it neigh\_state}[i]={\it
true}$ indicates that the node is adjacent to the node with the IP address $i$, while
${\it false}$ indicates that its adjacency status is either
\emph{unknown} or \emph{blacklisted} (since timing issues are not taken into account, these two statuses are considered the same). 

Lines 1-12 add the incoming route if no route previously exists. The route state of the incoming route is set in terms of the neighbor state of the sender message, i.e., ${\it neigh\_state}[{\it sip\_}]$. Lines 14-22 check whether the incoming route is a better route than the existing ones. If the incoming route is a  \emph{loop free}, in this solution a route which is not older or longer than the existing ones, then the routing table gets updated, lines 24-48. Lines 26-30 check whether there already exists a route from the sender that must be updated or it should be added to the first empty location. If the neighbor state of the sender is \emph{confirmed}, all the other routes must be cleared by reinitializing corresponding elements to $-1$, lines 37-43.

\subsection{Solution 2: Preferring Freshness to  Hop count}
In this approach, we favor incoming routes with greater sequence numbers over the existing routes even the valid ones. Since keeping routes with different sequence numbers jeopardizes the satisfaction of the loop freedom property, all the existing routes to the same destination as the incoming route must be removed from the routing table prior to adding the new route to the routing table. 
 \begin{itemize}
 	\item no route exists to the destination: the route is added to the routing table.
 	\item the sequence number of the incoming route is equal to the existing one while its cost, e.g. hop count, is equal or less than the existing one: the incoming route is added if the bound has not been reached already.
 	\item the sequence number of the incoming route is greater than all the existing routes to the same destination: the incoming route is added to the routing table after removing all the existing routes to the destination.
\end{itemize}
 The precise specification of this procedure is depicted in  \figurename{~\ref{code:UpdatingRoutingTable-solution2}}. Variables $\it rst$ and $\it nhops$ are defined two-dimensional similar to Section \ref{subsec::Solution1}. However, $\it dsn$ and $\it hops$ arrays are defined one-dimensional since in this solution we always keep routes with the greatest destination sequence number and the least hop counts, and hence all the route to the same destination will have the same destination sequence number and hop count. Lines 3-14 adds the incoming route to the routing table when no route exists to the destination. Then in line 18, the \emph{loop free} condition is checked, in this solution we consider a route loop-free if it has a larger destination sequence number or an equal one while it has a better hop count than the existing ones. If the neighbor state of the sender is \emph{confirmed}, the route must be added to the table with the \emph{valid} route state while all the other routes are cleared, lines 22-31. In lines 34-44, the routing table gets updated with the incoming route which has a better hop count. Otherwise, the route has a greater destination sequence number while its hop count is worse than the existing one. Lines 46-54 checks whether there exists a valid route to the destination. If there is no such a route to prevent loop formation, all the other routes must be cleared prior to adding the new route the routing table, Lines 56-66.

\begin{figure}
	\begin{center}
		\begin{lstlisting}[language=rebeca, multicols=2]
if(dsn[oip_]==-1)
{
	dsn[oip_]=osn_;
	if(neigh_state[sip_]==true)
	{
		rst[oip_][0]=1;
	}
	else
	{
		rst[oip_][0]=0;
	}
	hops[oip_]=hops_;
	nhop[oip_][0]=sip_;
	process_msg = true;
}
else
{
	if((dsn[oip_]==osn_ && hops[oip_]>=hops_) ||(dsn[oip_]<osn_ ))
	{
		if(neigh_state[sip_]==true)
		{
			rst[oip_][0]=1;
			for(int i=0;i<N;i++)
			{
				nhop[oip_][i]=-1; 
				dsn[oip_]=-1;
				hops[oip_]=-1;    
			} 
			hops[oip_]=hops_;
			nhop[oip_][0]=sip_;
			process_msg = true; 
		}
		else if(hops[oip_]>=hops_) {
			for(int i=0;i<N;i++)
			{
				if(nhop[oip_][i]==-1 || nhop[oip_][i]==sip_)
				{
					route_num = i;
					break;
				}
			}
			nhop[oip_][route_num]=sip_;
			rst[oip_][route_num]=0;
			process_msg = true; 
		} else if(dsn[oip_]<osn_){
			boolean vaild = false;
			for(int i=0;i<N;i++){
				if(rst[oip_][i]==-1){
				break;
				} else if(rst[oip_][i]==1){
					vaild = true;
					break;
				}
			}
			if(vaild == false){
				for(int i=0;i<N;i++)
				{
					nhop[oip_][i]=-1; 
					dsn[oip_]=-1;
					hops[oip_]=-1;   
				} 
				hops[oip_]=hops_;
				nhop[oip_][0]=sip_;
				rst[oip_][0]=0;
				dsn[oip_]=osn_;
				process_msg = true; 
			}
		}
	}                              
}
		\end{lstlisting}
	\end{center}
	\vspace{-2mm}
	\caption{The second solution for updating the routing table
		\label{code:UpdatingRoutingTable-solution2}}
\end{figure}

    \section{Related Work} \label{sec::related}
AODV as a routing protocol of MANETs, which is rapidly growing, drew lots of attention to itself. 
Modeling and verification of AODV‌ protocol has been the main topic of a great deal of studies. Many publications examined loop freedom as the most important property of this protocol through different approaches from extending the existing formal frameworks like SPIN~\cite{de2004formal,wibling2004automatized} and
UPPAAL~\cite{fehnker2012automated,mclver2006formal,wibling2005ad} to proposing new frameworks like CBS\#
\cite{nanz2006framework}, CWS \cite{Mezzetti2006331}, CMN
\cite{Merro2009194}, the $\omega$-calculus \cite{Singh2010440},
bA$\pi$ \cite{godskesen2010observables}, CMAN
\cite{godskesen2007calculus,Godskesen09}, RBPT
\cite{ghassemi2008restricted} and the bpsi-calculi
\cite{borgstrom2013broadcast,psi} to support the requirements of the new environment, i.e. MANETs, such as modeling the underlying topology, mobility and
local broadcast. However, these approaches can not be easily adopted by a user not familiar with  formal modeling concepts such as process algebra and timed automata. The Java-like syntax of wRebeca and its inherit friendliness brought up by the actor model, make it a suitable modeling approach that can be used by protocol designers at the early stages of their protocol development.

AODV was analyzed in \cite{wibling2004automatized} for some special mobility scenarios (as a part of the specification). A scenario leading to a loop was first discovered in \cite{spin3}. In \cite{fehnker2013process}, the route discovery procedure of AODV was  analyzed and it was shown that in \cite{vanGlabbeek2016} that for all arbitrary number of nodes, the protocol is loop-free. The loop freedom of AODVv2-04
for an arbitrary number of nodes was examined in \cite{LoopNamj} through an inductive and compositional proof: It provides an inductive invariant and proves that it is held initially and also preserved by every action, either a protocol action or a change in the network, similar to the approach of \cite{vanGlabbeek2016}. They have reported two loop-formation scenarios due to inappropriate setting of timing constants and accepting any valid route when the current route is broken without any further evaluation (to ensure loop formation). 

\cite{Fehnker2012AWN} propose a process algebra for wireless mesh networks, called AWN, which addresses the main challenges of MANETs. It demonstrates the applicability of such framework through model AODV and proving loop freedom condition. There are several studies such as \cite{HofnerRigorous, vanGlabbeek2016,Dymo,vanGlabbeekSequenceNumbers} that use AWN to model and analyze different versions of AODV and verify key properties of protocol namely loop freedom. \cite{Dymo} models dynamic MANET on-demand
(DYMO) routing protocol (also known as AODVv2) and shows how it solves some problems discovered in AODV and how it fails to address all the shortcomings.
\cite{HofnerRigorous} points out some ambiguities in the RFC then analyzes different readings of the AODV RFC, and show which interpretations are loop free. AWN is also used in \cite{vanGlabbeekSequenceNumbers} to shown that ambiguities in RFC can lead to loop formation and monotonically increasing sequence numbers, by
themselves, do not guarantee loop freedom. 

    \section{Conclusion and Future Work}
Aodv is a well-known and yet complex routing protocol which its most important property is loop freedom. Many studies showed the violation of this property over different version or even proved its correctness for some versions. Though loop freedom is preserved for some versions, an even a small change led to a loop formation. Therefore, it is desirable to have an ongoing verification in parallel with its design and development. In this paper, we illustrated the loop freedom violation for AODVv2-11,13 and 16 through $3$ counterexamples  and explained the reasons led to these loop formation scenarios. As the protocol evolves, counterexamples get more complex and harder to guess. Therefore, we need an automated tool to facilities the verification. wRebeca not only makes verification very easy by handling all difficulties of MANETs behind the scene, it results in an accurate specification. Having a precise specification prevents any ambiguity and facilitates the implementation. 
In addition to reporting the counterexamples, we proposed two solutions to make AODVv2-16 loop free, respecting two different aspects of the performance. One aspect favors new incoming routes over existing routes even the valid ones while the other  values validity over freshness. Then we showed how AODVv2-16 fails to recognize and use loop free routes and therefore fails to deliver a packet even if a loop free route does exist. This problem occurs due to over limitation of route maintenance. Finally, we proposed two solutions which are loop free while having better performance. They targeted two different aspects of performance as solutions for AODVv2-16. Based on our verifications for a network of five nodes while considering all possible topologies, by  not applying any network constraint, these two solutions are loop free. In addition, the proposed protocols are specified in a wRebeca model which has a Java-like syntax which makes it easy to read and comprehend.   

We plan to extend our framework to support timed aspects of MANET protocols to analyze real-time 
behavior of wireless network protocols. This extension enables us to model and verify the AODV protocol while considering its timing parameters.

     \bibliographystyle{elsarticle-num}
    \bibliography{ref}

\end{document}